\documentclass[a4paper,fleqn]{cas-dc}

\usepackage[numbers,sort&compress]{natbib}
\usepackage{subfig}
\usepackage{hyperref}
\usepackage{amsmath}
\RequirePackage[capitalise,nameinlink]{cleveref}
\usepackage{graphicx}
\usepackage{booktabs}
\usepackage{tabularx,ragged2e,siunitx}
\usepackage{xcolor,soul}
\usepackage[utf8]{inputenc}
\usepackage[normalem]{ulem}
\usepackage{amsmath}
\usepackage{float}

\def\tsc#1{\csdef{#1}{\textsc{\lowercase{#1}}\xspace}}
\tsc{WGM}
\tsc{QE}
\tsc{EP}
\tsc{PMS}
\tsc{BEC}
\tsc{DE}

\begin{document}
\let\WriteBookmarks\relax
\def\floatpagepagefraction{1}
\def\textpagefraction{.001}

\shorttitle{N-functionalized Non TiB}


\title [mode = title]{Functionalization-Driven Charge Redistribution Enabling Ultra-High-Capacity V$_2$B MBene Anode for  Li/Na ion batteries: A First-Principles study}




%
\author[1]{{Shaiokh Bin Abi}}





\credit{Conceptualization, Methodology, Visualization, Software, Investigation, Writing – original draft}

\affiliation[1]{organization={Department of Electrical and Electronic Engineering, Bangladesh University of Engineering and Technology},
    city={Dhaka},
    country={Bangladesh}}

\author[1]{Ahmed Zubair}[orcid=0000-0002-1833-2244]
\cormark[1]


\credit{Conceptualization, Methodology, Visualization, Resources, Writing – original draft, Writing – review \& editing, Supervision}




\cortext[cor1]{Corresponding author. email: ahmedzubair@eee.buet.ac.bd}




\begin{abstract}
Amid increasing global demand for clean, sustainable energy, the search for novel electrode materials has emerged as a crucial link to advancing future energy storage technologies. Here, we explored the potential of N-functionalized 2D MBene V$_2$BN$_2$ as anode materials for Li- and Na-ion batteries using first-principles calculations. Phonon dispersion and \textit{ab initio} molecular dynamics calculations were employed to assess the dynamic and thermal stability of the material. The intrinsic metallic properties of V$_2$BN$_2$ were revealed through electronic band structures and density of states analyses. Importantly, Bader charge analysis demonstrates substantial charge redistribution upon Li/Na adsorption, leading to stronger ion–substrate interactions compared to the pristine counterpart. This redistribution plays a decisive role in enhancing Li/Na ion adsorption and stabilizing ion accommodation. Furthermore, owing to favorable multilayer adsorption of Li and Na ions, V$_2$BN$_2$ exhibited high theoretical specific capacities of 1524 and 762 mAh/g, as well as low open circuit voltages of 0.73 and 0.23 V for Li and Na, respectively. In addition, the energy barriers were calculated to be 0.49 and 0.29 eV for Li- and Na-ion transport, respectively, indicating rapid ion transport and excellent rate capability. These results indicate that V$_2$BN$_2$ holds significant potential as an anode material for next-generation rechargeable ion batteries.

\end{abstract}







\maketitle

\section{Introduction}
  
For applications as electrodes in rechargeable ion batteries, considerable attention has been devoted to titanium-based anode materials. However, their limitations have underscored the need for alternative electrode candidates that deliver higher capacity, faster ion transport, and improved long-term stability in next-generation rechargeable batteries. This necessity has further been intensified by major developments in the energy sector over the past few decades. The advancement of rechargeable batteries with high energy and power densities, rapid charging and discharging rates, extended life cycles, and cost effectiveness would facilitate a gradual transition towards environmentally sustainable energy systems in the near future \cite{Good14EES,Gur18EES,Stoppato17IES}. Rechargeable batteries have been an integral part of not only large-scale energy storage systems but also in portable electronic devices and electric vehicles \cite{Lennon19MRS,Liu22ER}. Among these, lithium-ion batteries (LIBs) have dominated the market due to their high energy density, long cycle life, and excellent operational stability \cite{Asamoah24S,Adebanjo25EA}. While the practical applications of these batteries continue to grow, their structural degradation and inherent thermal instability persist as critical challenges to operational safety. In this context, sodium-ion batteries (SIBs) have emerged as promising candidates because of the natural abundance and low cost of sodium. Additionally, LIBs and SIBs have many similarities, including material components, the manufacturing methodologies of the materials and cells, and energy storage mechanisms. \cite{Van20CR,Gao24CE}. 

As a fundamental component of rechargeable ion batteries, the anode serves as the primary host for ion intercalation and deintercalation throughout charge–discharge cycles. Thus, the structure and properties of the anode material play a decisive role in governing the electrochemical performance and durability of the battery \cite{Nzereogu22ASSA}. Graphite anodes used in commercial LIBs offer a low theoretical specific capacity of 372 mAh/g \cite{Hassoun14NL,Xie17JC},  prompting an extensive search for suitable anode materials. In contrast, due to the larger atomic radius of the sodium ion, SIBs often exhibit inferior performance when employing the same electrode materials as LIBs. This limitation necessitates a deeper understanding of the underlying electrochemical mechanisms and the development of novel anode materials optimized for sodium storage. 

Since the advent of graphene, two-dimensional (2D) materials have gained significant research interest in the field of flexible electronics, photonics, and energy storage devices \cite{Xie17JC,Kim15ARMR,Lopez13NT}. Owing to inherent properties such as high surface area, enhanced electronic mobility, abundant active sites, tunable interlayer spacing, and structural flexibility, offer unique advantages compared with bulk nanomaterials \cite{Pumera25MC,Lin20CMS,Zhang17JPD}. For SIBs and LIBs, these features can help address key performance parameters: high capacity, rapid charge/discharge, and structural stability during repeated ion insertion and extraction. As promising anode materials for rechargeable ion batteries, the family of 2D materials has undergone remarkable expansion through a comprehensive research effort. This rapidly growing class includes a wide range of representatives such as hexagonal boron-nitride, transition metal dichalcogenides, transition metal oxides, silicene, germanene, borophene, phosphorene, and transition metal carbides/nitrides (MXenes).

Among 2D materials, transition metal borides (MBenes), boron analogs of MXenes, have recently emerged as a new class of layered compounds that exhibit exceptional electrical conductivity and strong mechanical rigidity. Furthermore, variable modes of 2D layer sandwiching \cite{Ade15IC} and distinct structural phases (hexagonal and orthorhombic) of MBenes offer an even wider tunability space than that of MXenes. Theoretical calculations performed on Mo$_2$B$_2$, Fe$_2$B$_2$\,\cite{Guo17JMC}, Ti$_2$B, and Sc$_2$B\,\cite{Ma22ASS}, demonstrated their promising potential as anode materials for rechargeable ion batteries. Experimentally synthesized MBenes such as MoB illustrated a high electrochemical performance, exhibiting a reversible specific capacity of 144.2 mAh/g after 1000 cycles at the current density of 2 A/g, surpassing the performance of many reported MXene-based anodes \cite{Xiong22CEJ}. Moreover, it showed excellent reversibility in SIBs \cite{Xiong25JMS}.

In recent times, the study of MXenes has been predominantly centered on titanium-based MXenes (Ti-MXenes) \cite{Lamiel23MT}, establishing the fundamental understanding of their tunable properties across diverse applications. Nevertheless, several other transition-metal systems remain underexplored and hold significant potential to expand the functions of 2D MXenes well beyond their Ti-based counterparts. In this context, vanadium—an earth-abundant transition metal—offers a promising alternative. V-based MXenes have increasingly attracted attention, with a growing number of experimental studies demonstrating their usefulness across a diverse range of fields such as electrocatalysis \cite{Wang23EcoMat} and electrochemical storage \cite{Vahid21ESM}. Furthermore, M$_2$B (Sc, Ti, and V), has emerged as a potential candidate for rechargeable magnesium ion batteries \cite{Ma22ASS}. Despite these advances, surface functionalization of V$_2$B, a relatively new MBene, remains insufficiently investigated. Functionalization of 2D materials plays a crucial role in tailoring their properties, making them more suitable as anode materials. It can improve the affinity between the anode surface and the alkali ions, facilitating efficient ion adsorption. Surface functionalization could modify the atomic structure and charge distribution, leading to improved ion diffusion kinetics. Furthermore, it can increase the number of available storage sites for foreign ions and thus effectively improve the battery capacity \cite{Khos20CAJ,Zheng22AM}. The pursuit of novel anode materials is driven by the demand for high energy density, rapid charging-discharging rate, sustainable material sourcing, and extended battery lifespan. While functionalized 2D MBenes have been reported as promising anode materials \cite{Liang22ASS,Xiao21ASS,Hu21CMS,Abi25ASS}, the electrochemical performance of functionalized V-based MBenes for Li- and Na-ion batteries remains largely unexplored.

In this work, density function theory (DFT) based first-principles calculations were performed to systematically investigate the functionalization of V$_2$B with relatively light group-V elements, namely nitrogen (N) and phosphorus (P), aiming to evaluate their feasibility as a potential candidate for anode materials in LIBs and SIBs. N and P was added to different sites on V$_2$B and the most stable structure was used for later calculations. Phonon dispersion and \textit{ab initio} molecular dynamics (AIMD) were performed to evaluate the thermodynamic stability of V$_2$BN$_2$ and V$_2$BP$_2$. Subsequently, a systematic analysis was carried out on the structures with Li/Na being adsorbed on them. Since an ideal rechargeable ion battery should exhibit high gravimetric and volumetric energy densities along with excellent rate capability, these performance metrics were thoroughly examined. The gravimetric and volumetric energy density are governed by the average voltage of the battery and the capacity, while the rate capability is governed by the kinetics of the ion transport. Hence, open circuit voltage (OCV), theoretical specific capacity, and diffusion barrier were calculated to assess the suitability of these materials as potential anode candidates.

\section{Computational Details}

All the electronic and thermodynamic computations were performed using first-principles calculations based on DFT by Quantum Espresso \cite{Giannozzi17JP}. The ion-electron interaction was described using the projected augmented wave (PAW) approach. Additionally, the Perdew-Burke-Ernzerhof (PBE) functional within the generalized-gradient approximation (GGA) was used for the exchange–correlation functional. The kinetic energy cutoff for wavefunctions and charge density were set to 60 Ry and 600 Ry, respectively. For simulating the monolayer, a sufficiently large vacuum of 20 \AA~ was introduced along the z-axis to avoid interlayer interactions. The geometries were optimized with a force tolerance of 10\textsuperscript{-3} Ry/Bohr and an energy tolerance of 10\textsuperscript{-5} Ry. The Brillouin zones of the primitive unit cell were sampled using Monkhorst–Pack meshes on (20$\times$20$\times$1) k-grid. The unit cells were functionalized using nitrogen and phosphorus at three possible sites. The most stable structures (with the lowest adsorption energies) were used for further calculations.

The unit cell was then repeated along the x and y directions to form a (2$\times$2$\times$1) supercell. In case of the supercell, self-consistent field (SCF) calculations were performed using a k-grid of (6$\times$6$\times$1), while non-self-consistent field (NSCF) calculations employed a denser mesh of (12$\times$12$\times$1). The band structure was evaluated along the ($\Gamma$-M-K-$\Gamma$) path of the first Brillouin zone, employing a uniform sampling of 20 k-points between successive high-symmetry points. Moreover, both the dynamical and energetic stabilities of the optimized structures were thoroughly examined. To assess the dynamic stability, phonon dispersion calculations were carried out using the finite-displacement supercell approach integrated within the PHONOPY framework \cite{Togo23JP}. The phonon calculations were performed using a k-grid size of (4$\times$4$\times$4). Furthermore, AIMD simulations were performed using the canonical (NVT) ensemble for evaluating thermal stability. The simulations were carried out at a temperature of 400 K for a duration of 5 ps with a time step of 1 fs. 

Subsequently, single Li and Na atoms were adsorbed on the supercell at three different sites. Along with average adsorption energy, Bader charge analysis\,\cite{Sanville07JCC} was employed to quantify the charge transfer occurring during Li/Na adsorption on the monolayer. The diffusion energy barriers for ion migration were computed using the climbing-image nudged elastic band (CI-NEB) method\,\cite{Henkelman00JCP}, employing linear interpolation with five intermediate images between two stable adsorption sites along each pathway.

 \begin{figure*}
    \centering
    \includegraphics[trim= {0cm 0cm 0cm 0cm},clip, width=\textwidth]{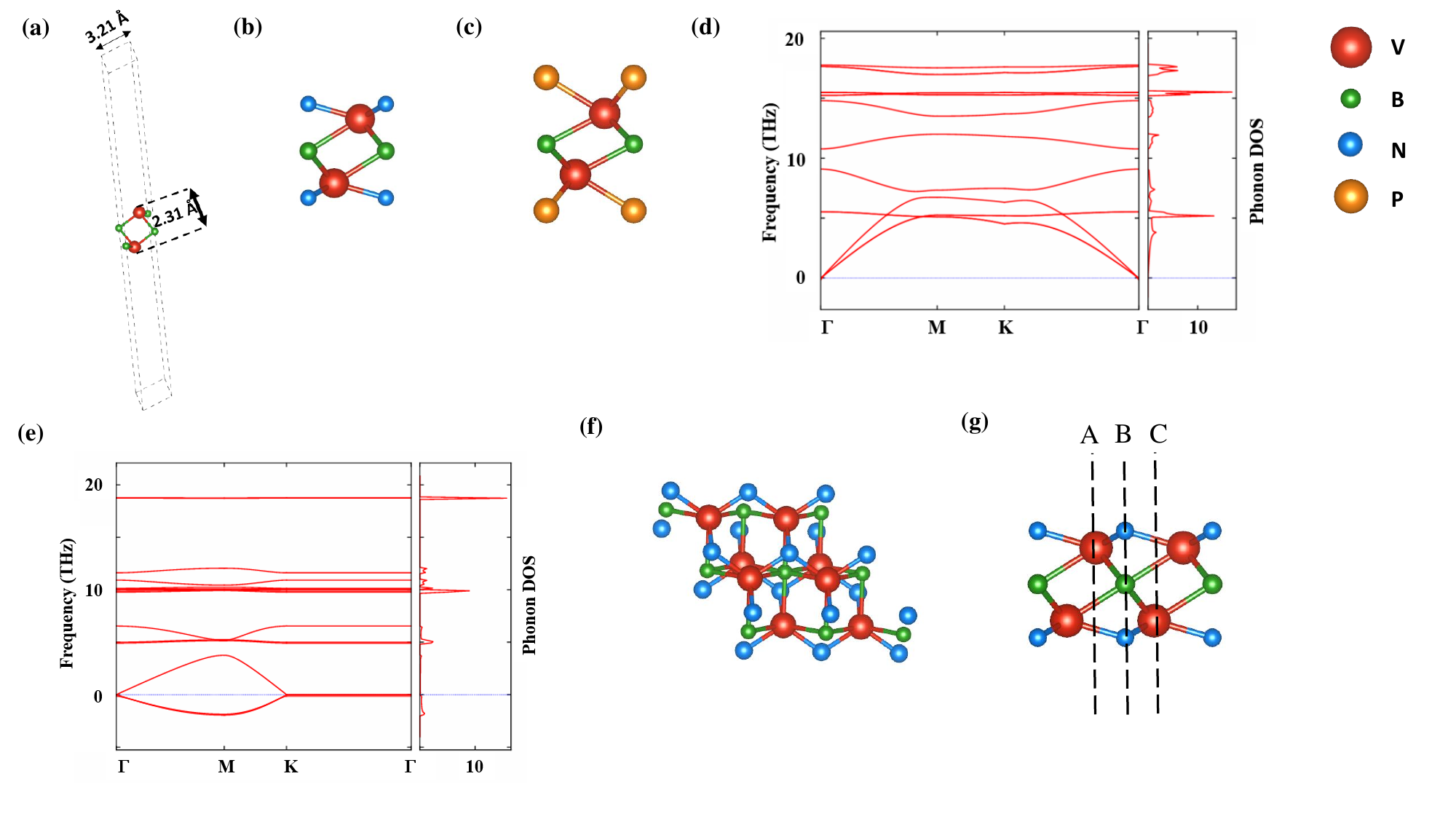}
    \caption{(a) Optimized unit cell of V$_2$B, (b) optimized unit cell of V$_2$BN$_2$, and (c) optimized unit cell of V$_2$PN$_2$. Phonon spectra and phonon density of states for unit cells of (d) V$_2$BN$_2$ and (e) V$_2$BP$_2$. (f) The perspective view of the optimized supercell of V$_2$BN$_2$, and (g) possible adsorption sites of a single Li or Na atom on V$_2$BN$_2$. }
    \label{fig:Fig.1}
\end{figure*}

The next step involved layer-by-layer adsorption of Li/Na atoms on the monolayer. For all simulations, van der Waals interaction with the Grimme type was considered at the DFT-D3 level \cite{Grimme10JCP}. The layers were added both above and below the monolayer, followed by geometric optimization. The process was repeated until the average adsorption energy turned positive. The corresponding maximum stable configurations were then employed to determine the theoretical specific capacities for Li and Na. The convex hull was plotted from the set of minimum-energy structures, and the open-circuit voltage (OCV) was subsequently calculated based on those stable states. To further assess the thermodynamic stability of the final structures, AIMD simulations were carried out for a total of 2 ps with a time step of 1 fs at a temperature of 350 K.

\section{Results and Discussion}

\subsection{Structural Properties and Thermodynamic Stability}

The optimized geometrical structure of V$_2$B is shown in Fig.~\ref{fig:Fig.1}(a). The unit cell consists of two vanadium atoms and one boron atom. It is a hexagonal lattice with a lattice constant a=b=3.21 \AA. The V-B bond length, the V-V bond length, and the V-B-V bond angle were found to be 2.18 \AA, 2.96 \AA~and 85.48$ ^{\circ}$. In addition, the buckling height was 2.31 \AA. These values are in close agreement with previously reported ones \cite{Ma22ASS}. The band and density of states (DOS) structures (see Fig. S1 of Supplementary Material) exhibit that V$_2$B is metallic. Additionally, it possesses a total magnetization of 1.12 $\mu_B$ per cell. Considering the structure's symmetry, three adsorption sites were chosen: on the top of the upper vanadium atom, on the top of the bottom vanadium atom, and on the top of the boron atom. At each of the three sites, two nitrogen (N) and phosphorus (P) atoms positioned above and below were introduced, and the resulting structures were subsequently relaxed. The binding energies for each of the configurations were calculated using the formula, 

\begin{equation}
    \mathrm{E_{b} = E_{V_{2}BX_{2}} - E_{V_{2}B}- 2E_{X}}. 
\end{equation}

In this expression, E$\mathrm{_{V_{2}BX_{2}}}$ and E$\mathrm{_{V_{2}B}}$ represents the energy of functionalized V$_2$BX$_2$ (X=N, P) and the energy of pristine V$_2$B, respectively. E$\mathrm{_X}$ denotes the energy of a single functionalization atom calculated using the same cell parameters as those of the V$_2$BX$_2$ structure. Negative values correspond to energetic feasibility. The structures with the lowest binding energies are shown in Figs.~\ref{fig:Fig.1}(b) and (c).


The next step involved investigating the physical stability. Thus, phonon simulations were performed on each of the unit cells of V$_2$BX$_2$ (X=N, P). The results are displayed in Fig.~\ref{fig:Fig.1}(d,e). As shown in Fig.~\ref{fig:Fig.1}(d), V$_2$BN$_2$ exhibits no imaginary frequencies, confirming its dynamic stability. In contrast, V$_2$BP$_2$ displays negative frequencies as seen in Fig.~\ref{fig:Fig.1}(e), indicating dynamic instability. Hence, V$_2$BN$_2$ was used for later calculations. 

\begin{figure*}
    \centering
    \includegraphics[trim= {0cm 5.5cm 0.5cm 1.5cm},clip, width=\textwidth]{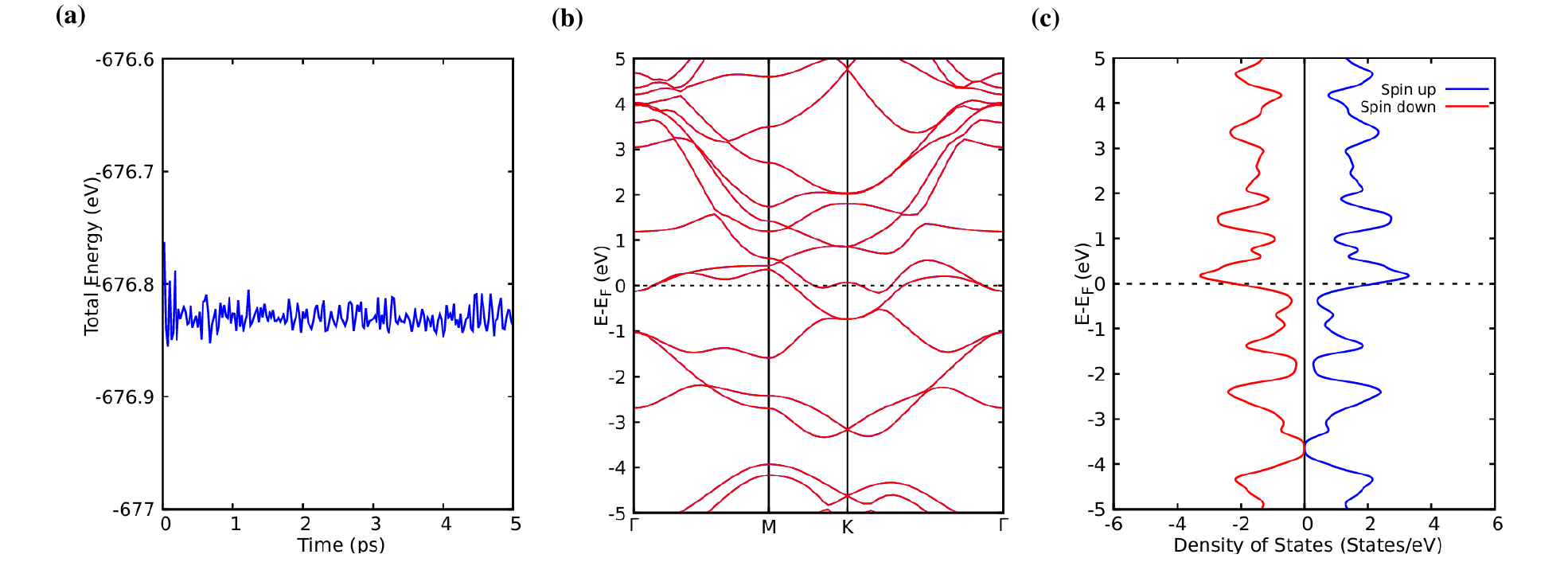}
    \caption{(a)The variation of energy at 400K with a duration of 5 ps by AIMD simulation, (b) electronic band structure and (c) density of states for V$_2$BN$_2$.}
    \label{fig:Fig.3}
\end{figure*}

Furthermore, the thermodynamic stability of V$_2$BN$_2$ was investigated using AIMD simulations. The calculation was performed at a temperature of 400 K, for a period of 5 ps, and with a time step of 1 fs. The result, in Fig.~\ref{fig:Fig.3}(a), shows that there are minimal energy fluctuations, which verifies the excellent stability of the mentioned structure. Fig.~\ref{fig:Fig.3}(b) and (c) illustrate the band structure and DOS of V$_2$BN$_2$ unit cell, respectively. The overlapping bands near the Fermi level demonstrate the metallic nature of V$_2$BN$_2$. Furthermore, the spin-up and spin-down plots were identical, which indicates the non-magnetic nature of the material, in contrast to pristine V$_2$B, which has finite magnetization. This non-magnetic nature is beneficial for anode materials, as it ensures uniform charge distribution and minimizes spin-related scattering, thereby enhancing electronic conductivity. 

\subsection{Adsorption Properties of Li/Na ion}

To investigate the suitability of V$_2$BN$_2$ as an anode material for Li / Na-ion batteries, the adsorption properties of Li/Na-ion on the V$_2$BN$_2$ monolayer were comprehensively studied. For this purpose, it was necessary to identify the most stable adsorption site. Thus, three adsorption sites were chosen as shown in Fig.~\ref{fig:Fig.1}(g). For convenience, the three sites are named Site A, Site B, and Site C, respectively. Li and Na were adsorbed on each of these sites. The adsorption energies were calculated by.

\begin{equation}
    \mathrm{E_{ad} = E_{V_{2}BN_{2}M} - E_{V_{2}BN_{2}}- E_{M}},
\end{equation}

where E$\mathrm{_{V_{2}BN_{2}M}}$ and E$\mathrm{_{V_{2}BN_{2}}}$ represent the energy of the system with and without adsorption of a Li or Na atom, respectively. E$\mathrm{_{M}}$ represents the energy of a single Li or Na atom using the same cell parameters as that of V$_2$BN$_2$ supercell. The adsorption energies for Li and Na at each site are presented in Table\,\ref{tbl1}.



\begin{table}[width=.9\linewidth,cols=6,pos=ht]
\caption{Adsorption energies of a single Li and Na atom at different sites of V$_2$BN$_2$ }\label{tbl1}
\begin{tabular*}{\tblwidth}{@{} LLL@{} }
\toprule
 Site & \multicolumn{2}{c}{E$_{ad}$(eV)}  \\
 & Li & Na  \\
\midrule

Site A &	--1.286 & --1.096   \\
Site B & --1.630 & --1.131  \\
Site C & --2.125  & --1.420  \\
\midrule
\end{tabular*}
\end{table}

Table \ref{tbl1} shows that Site C is the most suitable adsorption site for both Li and Na. The adsorption energies for Li are slightly more negative than those for Na, indicating a stronger interaction of Li with the monolayer compared to Na. This difference arises due to the higher electropositivity of Na compared to that of Li. Among the three evaluated sites, site A has the highest value, exhibiting the least favorable adsorption energy.  
 
To facilitate a better understanding of the adsorption of Li/Na on V$_2$BN$_2$ monolayer, Bader charge analysis and charge density difference plots were performed. Bader charge analysis revealed the average charge transferred between the atoms. The charge density difference plot was determined using the formula, 

\begin{equation}
    \mathrm{\Delta\rho = \rho_{V_{2}BN_{2}M} - \rho_{V_{2}BN_{2}}- \rho_{M}}. 
\end{equation}

In the above equation, $\rho$$\mathrm{_{V_{2}BN_{2}M}}$, $\rho$$\mathrm{_{V_{2}BN_{2}}}$ and $\rho$$\mathrm{_{M}}$ are the charge density of the monolayer including Li/Na atom, the monolayer and single atom, respectively.

\begin{table}[width=.9\linewidth,cols=4,pos=ht]
\caption{Average Bader charge transfer for adsorption of a  single Li/Na atom on V$_2$BN$_2$ }\label{tbl4}
\begin{tabular*}{\tblwidth}{@{} LLLL@{} }
\toprule
\multicolumn{4}{c}{Average Bader Charge Transfer (e)} \\
 Structures & Li/Na  & V & N\\
\midrule
V$_2$BN$_2$Li$_{0.25}$ & +0.873 & +1.386 & --1.579	 \\ 
V$_2$BN$_2$Na$_{0.25}$ & + 0.808 & +1.398 &	--1.556	 \\

\bottomrule
\end{tabular*}
\end{table}

The results of the Bader charge analysis are displayed in Table 2. The data reveals that for Li the value of charge transferred is +0.873, and the value for that of Na is +0.808. In comparison, adsorption of a single Li/Na atom on the pristine V$_2$B supercell results in relatively lower charge transfer, as summarized in Table S1 (see Supplementary Material). Specifically, the charge transferred amounts to +0.834 for Li and +0.640 for Na, which are both smaller than the corresponding values obtained for V$_2$BN$_2$. A notable difference is also observed in the charge contribution from vanadium atoms. For V$_2$BN$_2$, the charge transfer from V atoms is +1.386 and +1.398 for Li and Na adsorption, respectively, whereas for pristine V$_2$B, the corresponding values decrease to +0.496 and +0.574. These results clearly indicate that N-functionalization of V$_2$B induces significant charge redistribution, thereby enhancing the interaction strength between the adsorbed Li/Na ions and the substrate.

\begin{figure}[htbp]
    \centering
    \includegraphics[trim= {1cm 9cm 0cm 0cm},clip, width=9cm]{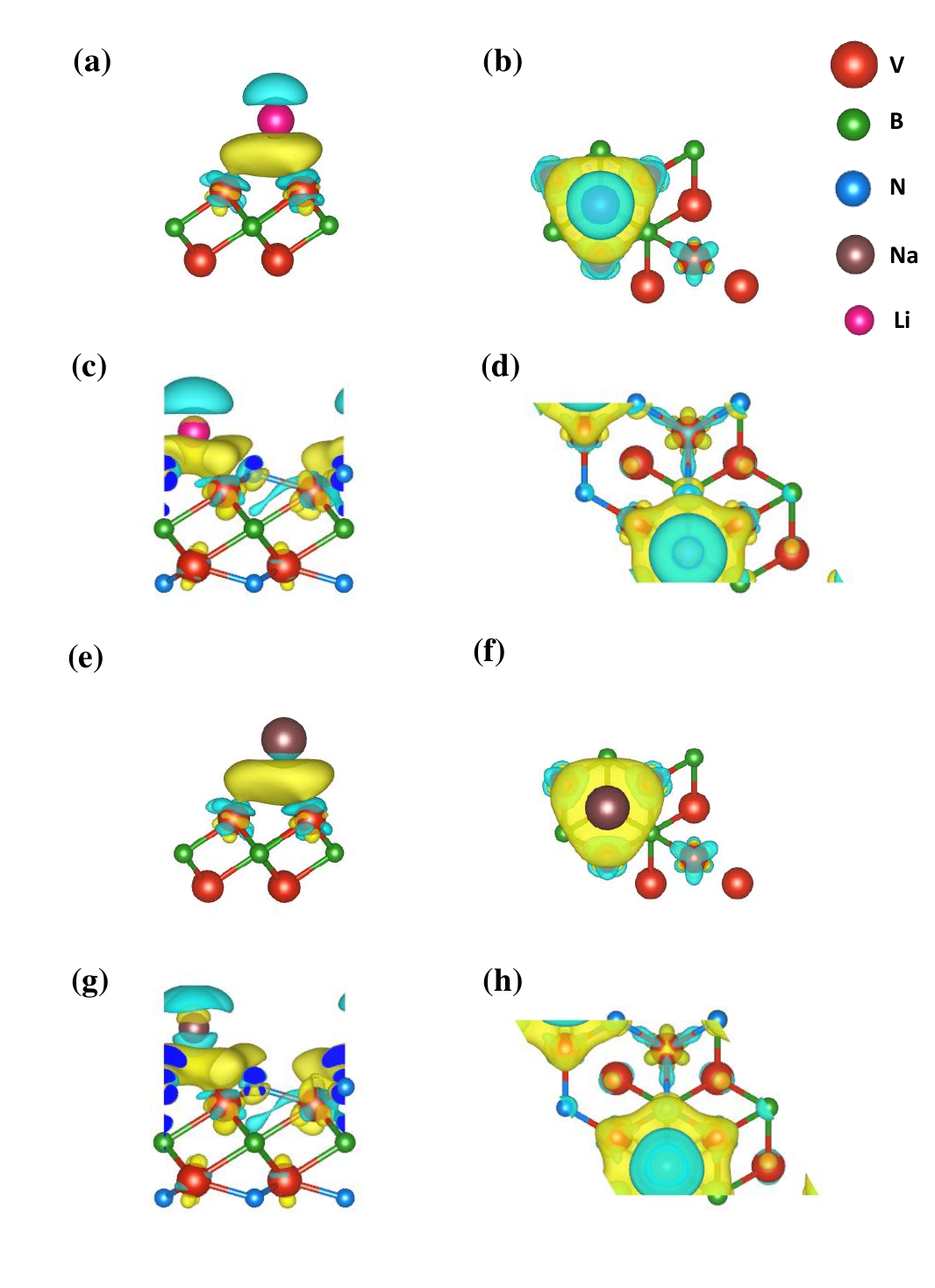}
    \caption{Charge density difference plots for adsorption of a single Li and Na atom, side view and top view, (a,b) V$_2$BLi$_{0.25}$ (c,d) V$_2$BN$_2$Li$_{0.25}$, (e,f) V$_2$BNa$_{0.25}$ and (g,h) V$_2$BN$_2$Na$_{0.25}$. }
    \label{fig:Fig.3a}
\end{figure}

The charge density plots for adsorption of Li and Na on pristine V$_2$B and V$_2$BN$_2$ are shown in Fig.~\ref{fig:Fig.3a} . The cyan colored region represents electron depletion, and the yellow color represents a region of electron accumulation. The isosurface level was set to 0.002 e {\AA}\textsuperscript{-3} and 0.0015 e {\AA}\textsuperscript{-3} for Li and Na adsorption, respectively. Cyan regions appear near Li/Na atom and the closest vanadium atom, whereas the yellow-colored region appears near the neighboring nitrogen atoms (for V$_2$BN$_2$). A key difference was that the charge transfer for Li adsorption on V$_2$B was slightly less than that in the case of V$_2$BN$_2$. However, in the case of Na adsorption, the charge transfer that occurs in V$_2$B monolayer was significantly increased in the case of V$_2$BN$_2$. The charge density difference analyses revealed that functionalization of V$_2$B with N leads to a stronger interaction between the Li/Na atom and the substrate. Thus, both Bader charge analysis and charge density difference plots indicate significant charge transfers between the atoms, which verifies the chemical adsorption of Li/Na atom on V$_2$BN$_2$ monolayer.

\subsection{Ion migration on V$_2$BN$_2$}

 \begin{figure*}[b]
    \centering
   \includegraphics[trim= {1cm 6cm 1cm 2cm},clip, width=\textwidth]{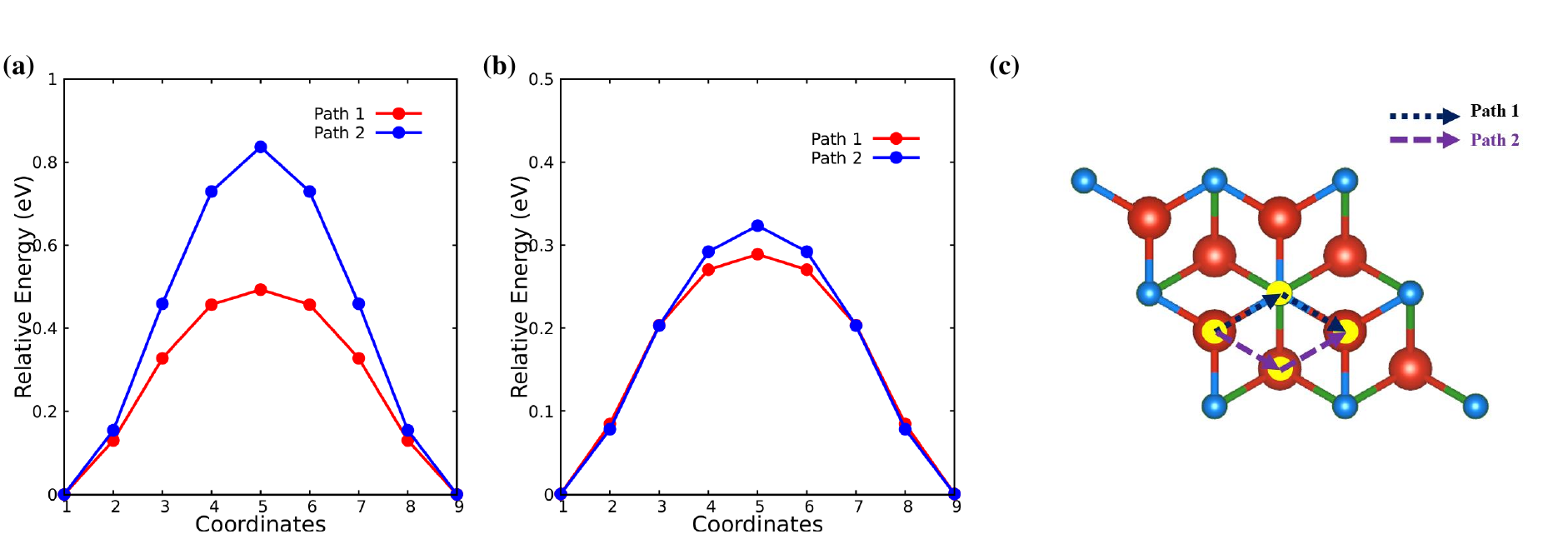}
 \caption{ Energy barrier profile for the diffusion of (a) Li and (b) Na on V$_2$BN$_2$.  (c) Possible paths for Li or Na diffusion.}
   \label{fig:enter-Fig.6}
\end{figure*}

\begin{figure*}[b]
    \centering
    \includegraphics[trim= {3cm 0cm 3cm 0cm},clip, width=\textwidth]{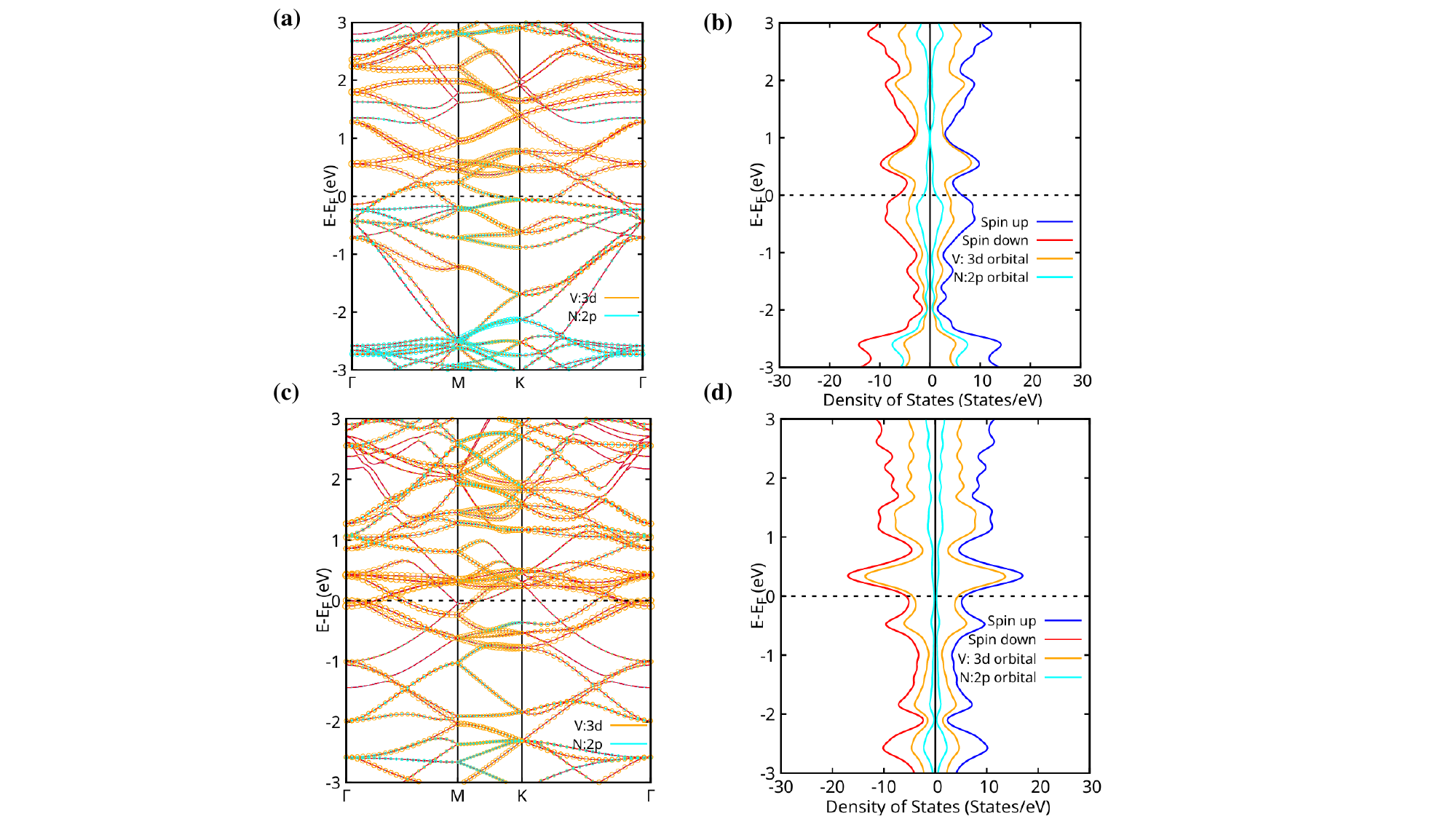}
  \caption{ Projected band structure and density of states for single layer of (a,b) Li and  (c,d) Na adsorbed on V$_2$BN$_2$.} 
    \label{fig:enter-Fig.7}
\end{figure*}

The diffusion mechanism of Li/Na ions is directly related to the charging and discharging rates of the battery. A low diffusion barrier enables the ions to move rapidly within the electrodes, thereby causing fast charging/discharging rates. In this study, the diffusion of Li/Na ions on V$_2$BN$_2$ monolayer was investigated using the CI-NEB method. The two paths considered for ion migration are shown in Fig.~\ref{fig:enter-Fig.6}(c). Path 1 started from site A, then to site C, and ended at another site A, whereas path 2 consisted of site A-site B-site C. For both these paths, nine images were considered for the CI-NEB method. The resulting energy profile for lithium is shown in Fig.~\ref{fig:enter-Fig.6}(a) and Fig.~\ref{fig:enter-Fig.6}(b) corresponds to that of sodium. For lithium, the diffusion energy barriers along path 1 and path 2 were calculated to be 0.49 and 0.84 eV, respectively, whereas for sodium, the corresponding values are 0.29 and 0.32 eV. In both systems, ion migration along path 1 is energetically more favorable than along path 2. While the barriers were comparable for sodium, a notable difference was observed for lithium. Furthermore, the diffusion energy barriers validate the suitability of V$_2$BN$_2$ as an anode for both Li and Na ion batteries. The comparatively lower energy for Na implied better ionic mobility.

\subsection{Improved Specific Capacity and Favorable OCV}

In addition to the diffusion barrier, further investigations were carried out to elucidate the electrochemical properties of V$_2$BN$_2$ as a potential anode material. Accordingly, two important parameters -- OCV and theoretical specific capacity -- were systematically examined. For this purpose, layered Li and Na adsorption was performed on the supercell (2$\times$2$\times$1) of V$_2$BN$_2$.  As site C was the most favorable adsorption site for both Li and Na, four Li and Na atoms were adsorbed above and below V$_2$BN$_2$ to form the first layer. Afterward, the structure was optimized to obtain the most stable configuration, and the average adsorption energy for the first layer of ions was calculated by,

\begin{equation}
    \mathrm{{E^{1}}_{ads} =\frac{{E_{V_{2}BN_{2}M_{z}} - E_{V_{2}BN_{2}}- zE_{M}}}{z}},  
\end{equation} 

where, E$\mathrm{_{V_{2}BN_{2}M_{z}}}$ represents the energy of V$_2$BN$_2$ substrate adsorbed by z number of M atoms (M=Li, Na), E$ \mathrm{_{V_{2}BN_{2}}}$ represents the energy of substrate and E$ \mathrm{_{M}}$ represents the energy of single M atom.

The first-layer adsorption of Li and Na yielded negative average adsorption energies, indicating favorable adsorption. The electronic properties of the structure were further explored by computing the band structure and the projected density of states (PDOS). The results are depicted in Fig.~\ref{fig:enter-Fig.7}. In both cases, the spin-up and spin-down plots were identical indicating spin degeneracy. Thus, the non-magnetic nature and metallic characteristics were both retained after Li and Na adsorption. Moreover, the plots revealed that close to the Fermi level, significant contributions were made by 3d orbitals of vanadium atoms. Thus, these orbitals play a crucial role in governing the electronic conductivity of the system. Owing to this metallic conductivity, the suitability of V$_2$BN$_2$ as a potential anode material is further verified. 

Layered adsorption was continued to find the maximum number of Li/Na that could be accommodated on the monolayer. For the second layer, four Li/Na atoms were adsorbed above and below V$_2$BN$_2$ at site C. The average adsorption energies, for Li/Na, were calculated by,
\begin{equation}
    \mathrm{{E^{n}}_{ads} =\frac{{E_{V_{2}BN_{2}M_{n}} - E_{V_{2}BN_{2}M_{n-1}}- (M_{n}-M_{n-1})E_{M}}}{(M_{n}-M_{n-1})}}.  
\end{equation}

Here, n represents the number of Li or Na atomic layers, M$\mathrm{_n}$ and M$\mathrm{_{n-1}}$ are the number of Li or Na atoms in the n-th and (n-1)-th layer, E$\mathrm{_{V_{2}BN_{2}M_{n}}}$ and E$\mathrm{_{V_{2}BN_{2}M_{n-1}}}$ are the energies of the structures after adsorbing M$\mathrm{_n}$ and M$\mathrm{_{n-1}}$ Li or Na atoms. 

 For the second adsorption layer, Li/Na atoms were placed at Site B, which was identified as the second most energetically favorable site (see Table \ref{tbl1}). Consequently, the third layer was constructed by adsorbing atoms at Site A (the least favorable adsorption site). The process was repeated until the average adsorption energy became positive, signifying that further adsorption was energetically unfavorable. Table S1 in the Supplementary Material presents the calculated average adsorption energies for the layered configuration. The V$_2$BN$_2$ monolayer was found to accommodate up to four Li layers (above and below the surface), whereas only two layers could be stably held in the case of Na. Additionally, the change in lattice parameters after Li/Na adsorption are given in Table S2 in Supplementary Material. The mismatch for the fully lithiated and sodiated structures was found to be 0.51\% and 2.01\%, respectively. These values suggest that the monolayer can maintain its structural integrity during ion intercalation. Since theoretical specific capacity measures the number of Li/Na ions that using the maximum number of adsorbed Li/Na atoms, theoretical specific capacity is calculated by, 
 \begin{equation}
    \mathrm{C =\frac{AN_{max}F}{M_{V_{2}BN_{2}}}},
\end{equation}

where A stands for the valence number (A = 1 for Li and Na), N$\mathrm{_{max}}$ represents the maximum number of metal atoms that can be adsorbed, F denotes Faraday constant (26801 mAh/mol)  and  M$\mathrm{_{V_{2}BN_{2}}}$ is the molar mass of the structure. The specific capacities for Li and Na were calculated to be 1524 and 762 mAh/g, respectively. 
 
 \begin{figure*}[b]
    \centering
   \includegraphics[trim= {1cm 0cm 4cm 0cm},clip, width=\textwidth]{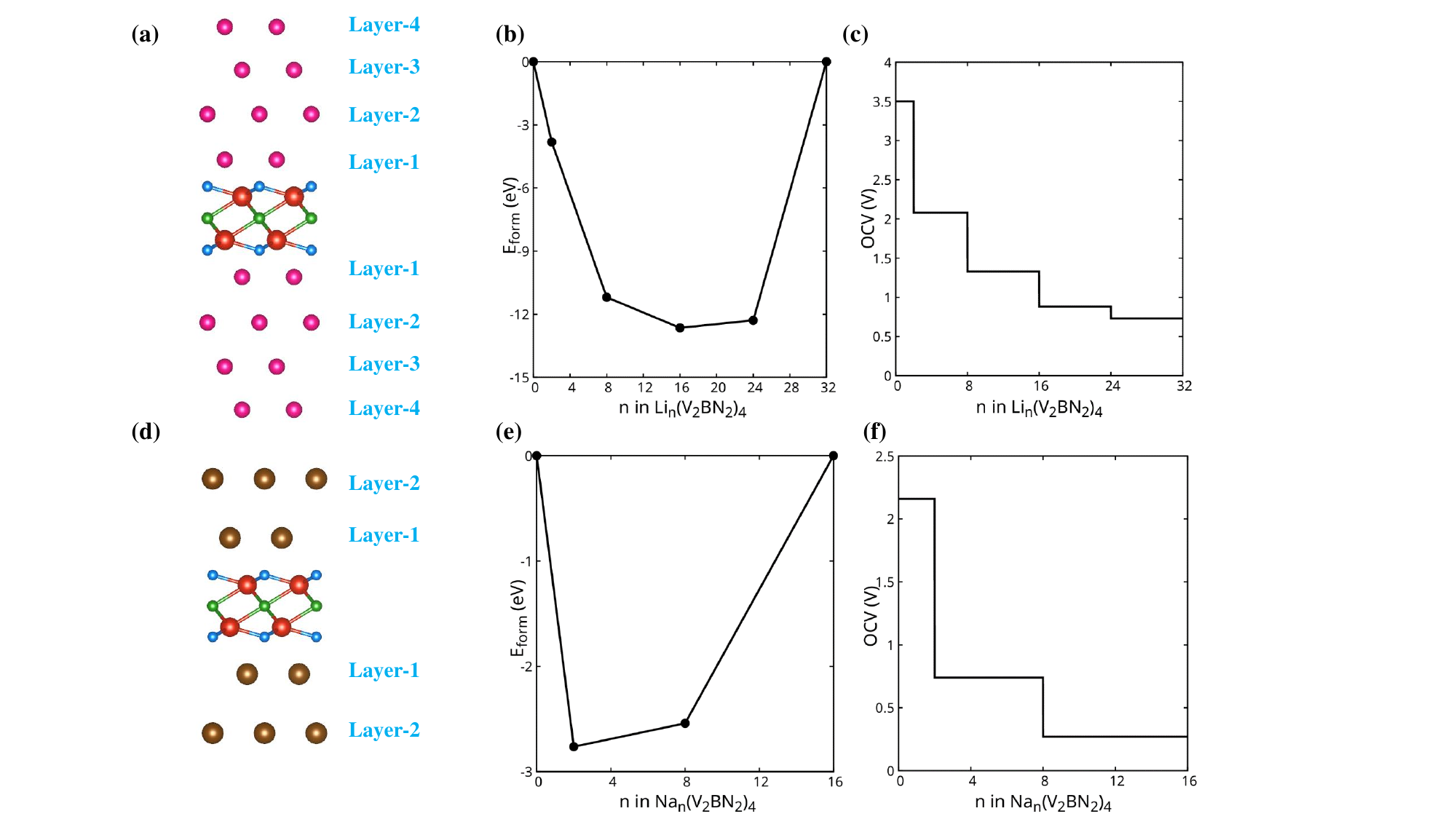}
 \caption{ (a) Configuration of maximum Li adsorbed on V$_2$BN$_2$, and (b) convex hull plot and (c) OCV for adsorption of Li on V$_2$BN$_2$. (d) Maximum Na adsorption configuration on V$_2$BN$_2$, and (e) convex hull diagram and (f) OCV profile for Na adsorption on V$_2$BN$_2$. }
   \label{fig:enter-Fig.8}
\end{figure*}

Another important parameter for evaluating and designing anode materials is OCV. To determine this, the formation energies were calculated using the following expression.
 
 \begin{equation}
  \begin{split}
    \mathrm{ E_{form} = E_{M_{N}(V_{2}BN_{2})_{p}} -\frac{{NE_{{M_{N_{max}}}(V_{2}BN_{2})_{4}}}} {N_{max}} } \\
   \mathrm{ - \frac{(N_{max}-N)E_{(V_{2}BN_{2})_{p}}} {N_{max}} } ,  
  \end{split} 
\end{equation}

here, p represents the number of unit cells (p=4), 
E$\mathrm{_{M_{N}(V_{2}BN_{2})_{p}}}$ and E$\mathrm{_{M_{N_{max}}(V_{2}BN_{2})_{p}}}$ represent the total energies of V$_2$BN$_2$ structures after adsorbing $ \mathrm{N}$ and $ \mathrm{N_{max}}$ M atoms, respectively. $\mathrm{N}$ represents the number of M atoms adsorbed, and $ \mathrm{N_{max}}$ represents the maximum number of M atoms that can be adsorbed on V$_2$BN$_2$ surface. And E$\mathrm{_{(V_{2}BN_{2})_{p}}}$ represents the total energy of V$_2$BN$_2$ supercell. Using the configurations that corresponded to minimum energy, the convex hull was plotted (Fig.~\ref{fig:enter-Fig.8}(a, c)). In case of Li adsorption, there were five intermediate stages (n=2,8,16,24,32) and for Na three intermediate stages (n=2,8,16) were found. Furthermore, using these stable phases, OCV values were calculated by,
 
 \begin{equation}
   \mathrm{ OCV = - \frac{{E_{V_{2}BN_{2}+q_{2}M} - E_{V_{2}BN_{2}+q_{1}M}- (q_{2}-q_{1})E_{M}}}{(q_{2}-q_{1})e}}. 
\end{equation}

In this expression, E$\mathrm{_{V_{2}BN_{2}+q_{2}M}}$ and E$\mathrm{_{V_{2}BN_{2}+q{1}M}}$ correspond to the total energies of V$_2$BN$_2$ structures after the adsorption of $ \mathrm{q_{2}}$ and $ \mathrm{q_{1}}$ number of M atoms, respectively, and E$ \mathrm{_M}$ represents the energy of a single M atom. The results are displayed in (Fig.~\ref{fig:enter-Fig.8}) (b, d). The Li system exhibited five voltage platforms, which decreases from 3.50 to 0.73 V, whereas for Na, OCV had three platforms whose value dropped from 2.16 to 0.27 V. Furthermore, thermal stability of the fully lithiated and sodiated strutures were examined using AIMD simulations. The simulations were performed for a total duration of 2 ps using a time step of 1 fs at a temperature of 350 K. The results, shown in Fig. S1 of the Supplementary Material, confirmed the thermodynamic stability of the structures.

Since OCV is a reflection of how strongly Li/Na ions bind to the anode material, if voltage is high, then it indicates strong ion binding, which in turn lowers the overall energy density, and if the voltage is low, it could lead to the growth of metal clusters. Hence, a suitable range for OCV is considered to be within 0.2-1 V\,\cite{Eames14JAC}. For V$_2$BN$_2$, the fully lithiated and sodiated structure demonstrated OCV values of 0.73 V and 0.27 V, respectively, indicating its suitability for Li/Na ion batteries. 

\begin{table*}[ht]
\caption{Performance comparison of V$_2$BN$_2$ with state-of-the-art anode materials for Li/Na-ion batteries. }
\label{tbl3}
\begin{tabular}{>{\centering\arraybackslash}p{2.5cm}
 >{\centering\arraybackslash}p{1.5cm}
 >{\centering\arraybackslash}p{3.8cm}
 >{\centering\arraybackslash}p{1.75cm}
 >{\centering\arraybackslash}p{3cm}
 >{\centering\arraybackslash}p{1.5cm}}

\hline

Type of metal ion battery & Material  & Specific Capacity (mAh/g) & OCV (V) & Diffusion Barrier (eV) & Reference  \\
\hline
\multirow{7}{*}{Li}
&Ti$_2$BN$_2$   & 398 & 0.93  & 0.44  &  \cite{Liang22ASS} \\ 
&Sr$_2$RuO$_2$ & 1211 & 0.48   & 0.38  & \cite{Martins25JPC} \\  
&Ti$_2$NbC$_2$O$_2$  & 274 & 0.81  & 0.30  & \cite{Perez25RSC} \\
&Mo$_2$B$_2$S$_2$  & 521 & 0.05  & 0.33  & \cite{Ahmad24NT} \\
&V$_2$BN$_2$  &  1524  &  0.73 &  0.49 & This work  \\ 

\hline
\multirow{7}{*}{Na}
&Ti$_2$BN$_2$    & 797 & 0.27  & 0.34  &  \cite{Liang22ASS} \\ 
&Ta$_2$Se$_2$C   & 404 & 0.26  & 0.11  & \cite{Martins25JPC} \\ 
&Ti$_2$NbC$_2$O$_2$  & 219 & 0.43  & 0.18  & \cite{Perez25RSC} \\
&Mo$_2$B$_2$S$_2$   & 298 & 0.36  & 0.37  & \cite{Ahmad24NT} \\
&V$_2$BN$_2$  & 762  &  0.27 &  0.29  &This work  \\ 
        
\hline
\end{tabular}
\end{table*}

\section{Performance Comparison }

In this section, a comparative analysis was conducted using the evaluated electrochemical properties- OCV, theoretical specific capacity and diffusion energy barrier against those of previously reported anode materials (Table \ref{tbl3}). Such a comparison provides valuable insights into how the performance of the proposed material compares to state-of-the-art candidates of anodes for LIBs and SIBs.

From the results shown in Table \ref{tbl3}, it is evident that the proposed material has a higher theoretical maximum capacity, for Li, than previously reported MXenes like Sr$_2$RuO$_2$ \cite{Martins25JPC} and Ti$_2$NbC$_2$O$_2$ \cite{Perez25RSC}. It is also higher than the values reported for other functionalized MBenes, including Ti$_2$BN$_2$ \cite{Liang22ASS} and Mo$_2$B$_2$S$_2$ \cite{Ahmad24NT}. Moreover, calculated OCV values for V$_2$BN$_2$ are lower than Ti$_2$BN$_2$ \cite{Liang22ASS} and Ti$_2$NbC$_2$O$_2$ \cite{Perez25RSC}. The value, in case of Li, lies within the suitable range of 0.2-1 V \cite{Eames14JAC}, indicating its suitability as an anode material and suggesting a reduced likelihood of dendrite formation during lithiation. Another important parameter that is critical to the performance of anode materials is the diffusion energy barrier. A low diffusion barrier exemplifies rapid ion kinetics, which enhances battery performance. Table \ref{tbl3} demonstrates that, for Li, the evaluated energy barrier in this study is comparable to some recently published MXenes and MBenes. 

In case of Na-ion, the specific capacity of V$_2$BN$_2$ (762 mAh/g) is substantially high, surpassing previously studied MXene and MBene systems such as Ta$_2$Se$_2$C \cite{Martins25JPC},  Ti$_2$NbC$_2$O$_2$ \cite{Perez25RSC}, and Mo$_2$B$_2$S$_2$ \cite{Ahmad24NT}. Its specific capacity is also comparable to Ti$_2$BN$_2$ \cite{Liang22ASS}, demonstrating its strong potential for high-energy-density applications. The OCV of V$_2$BN$_2$ lies within the convenient range for anode materials. This value is identical to that of Ti$_2$BN$_2$ \cite{Liang22ASS} and lower than the value mentioned for Ti$_2$NbC$_2$O$_2$ \cite{Perez25RSC}, indicating a high-energy configuration following sodium ion intercalation. In terms of ion transport kinetics, V$_2$BN$_2$ exhibits a diffusion barrier of 0.29 eV. This barrier is lower than those calculated for Ti$_2$BN$_2$ \cite{Liang22ASS} and Mo$_2$B$_2$S$_2$ \cite{Ahmad24NT}. On the other hand, MXenes such as Ta$_2$Se$_2$C \cite{Martins25JPC} and Ti$_2$NbC$_2$O$_2$ \cite{Perez25RSC} exhibited lower barriers. Overall, the low barrier in V$_2$BN$_2$ suggests efficient ion diffusion pathways that support excellent rate performance. Combining the high theoretical capacity, optimal OCV, and low diffusion barrier, the proposed material (V$_2$BN$_2$) demonstrates strong potential as a high-performance anode material for next-generation LIBs and SIBs.

\section{Conclusion}

In this paper, the potential of V$_2$BN$_2$ as an anode material for Li and Na-ion batteries was systematically investigated within the framework of DFT. V$_2$B was chosen as the parent MBene because  previous studies had predominantly focused on titanium-based MBenes, while  vanadium-based MBenes remained relatively unexplored despite their promising electrochemical characteristics. The thermal and dynamic stabilities of V$_2$BN$_2$ were verified. Subsequently, based on symmetry, three adsorption sites were selected to investigate the adsorption properties of single Li and Na ions on the monolayer. The calculated negative adsorption energies indicate strong ion–substrate interactions, which are further substantiated by Bader charge analysis revealing significant charge redistribution upon adsorption.
This redistribution plays a crucial role in enhancing the binding of Li/Na ions and facilitating efficient charge transfer between the adsorbates and the substrate. Moreover, the migration of Li and Na ions on V$_2$BN$_2$ surface was examined by calculating the diffusion energy barrier. The lowest energy barriers were found to be 0.49 and 0.29 eV for Li and Na, respectively, which are better than those of previously reported 2D materials. For further analysis, multilayer adsorption of Li and Na ions was performed on V$_2$BN$_2$. The band and DOS structures demonstrated metallic properties of the structures before and after adsorption, which is beneficial for efficient charge transport. Performance parameters such as specific capacity and OCV were found to be 1524 and 762 mAh/g, and 0.73 and 0.27 V for Li and Na, respectively. In addition, molecular dynamics calculations confirmed the thermal stability of fully loaded Li and Na structures. Overall, V$_2$BN$_2$ examined in this study provides intrinsic metallic characteristics, thermal and dynamic stability, low energy barrier and OCV, and high theoretical specific capacity,  highlighting this functionalized non-Ti based MBene as a promising anode material for rechargeable ion batteries.

\printcredits
\section*{Data Availability Statement}
{The data supporting the findings presented in this paper are not currently available to the public, but they may be obtained from the authors upon reasonable request.} 

\section*{Acknowledgements}
The authors thank the Bangladesh University of Engineering and Technology (BUET) for providing technical support. 

\bibliographystyle{model2-num-names}
\bibliography{cas-refs}

\end{document}